\documentclass{PoS}

\title{NectarCAM : a camera for the medium size telescopes of the Cherenkov Telescope Array}

\ShortTitle{NectarCAM}

\author{{J-F. Glicenstein$^9,$ O.~Abril$^6,$
 J-A.~Barrio$^{14}$, 
 O.~Blanch~Bigas$^{6}$, 
 J.~Bolmont$^{12}$, 
 F.~Bouyjou$^{9}$,
 P.~ Brun$^{13},$ 
E.~Chabanne$^{10},$ 
C.~Champion$^{1}$, 
S.~Colonges$^{1}$,
P.~Corona$^{12}$, 
E.~Delagnes$^9,$ 
C.~Delgado$^2$,
C.~Diaz Ginzo$^2,$ 
D.~Durand$^9,$ 
J-P.~Ernenwein$^3,$
S.~Fegan$^{11}$, 
O.~Ferreira$^{11}$, 
M.~Fesquet$^9,$ 
A.~Fiasson$^{10}$, 
G.~Fontaine$^{11},$ 
N.~Fouque$^{10},$ 
D.~Gascon$^{5},$ 
B.~Giebels$^{11},$ 
F.~Henault$^{7},$
R.~Hermel$^{10},$ 
D.~Hoffmann$^{3},$ 
D.~Horan$^{11},$ 
J.~Houles$^{3},$ 
P.~Jean$^{8},$ 
L.~Jocou$^{7}$,
S.~Karkar$^{12},$ 
J.~Kn\"odlseder$^{8},$ 
R.~Kossakowski$^{10},$ 
G.~Lamanna$^{10}, $
T.~Le~Flour$^{10}, $
J-P.~Lenain$^{12}, $
A.~Leveque$^{8},$ 
F.~Louis$^{9},$ 
G.~Martinez$^{3},$
Y.~Moudden$^{9},$ 
E.~Moulin$^{9},$
P.~Nayman$^{12},$
F.~Nunio$^{9},$
J-F.~Olive$^{8},$
J-L.~Panazol$^{10},$
S.~Pavy$^{11},$
P-O.~Petrucci$^{7},$
E.~Pierre$^{12},$
J.~Prast$^{10},$
M.~Punch$^{1}$, 
P.~Ramon$^{8},$
S.~Rateau$^{11},$
T.~Ravel$^{8}$,
S.~Rosier-Lees$^{10},$
A.~Sanuy$^{5},$
M.~Shayduk$^{4,9}$, 
P-Y~Sizun$^{10},$
K-H.~Sulanke$^{4},$
J-P.~Tavernet$^{12},$
L-A.~Tejedor~Alvarez$^{14},$
F.~Toussenel$^{12},$
G.~Vasileiadis$^{13},$
V.Voisin$^{12},$ 
V.~Waegebert$^{8},$
R.~Wischnewski$^{4}$,for the CTA consortium$^{15}$} \\
  {
  1: APC, Univ Paris Diderot, CNRS/IN2P3, CEA/lrfu, Obs de Paris, Sorbonne Paris Cit\'e, France, France,  \\
  2: CIEMAT, Spain,\\
  3:Centre de Physique des Particules de Marseille (CPPM), Aix-Marseille Universit\'e, CNRS/IN2P3, Marseille, France,\\
  4: Deutsches Elektronen-Synchrotron, Germany,\\
  5: Departament d'Astronomia i Meteorologia, Institut de Ci\`encies del Cosmos, Universitat de Barcelona, Spain, \\ 
  6: Institut de F\'isica d'Altes Energies, Spain,\\
  7: Institut de Plan\'etologie et d'Astrophysique de Grenoble, INSU/CNRS, Universit\'e Joseph Fourier, France,\\
  8: Institut de Recherche en Astrophysique et Plan\'etologie, France, \\
  9: IRFU, CEA-Saclay, Gif-sur-Yvette, France,\\
  10: Laboratoire d'Annecy-le-Vieux de Physique des Particules, Universit\'e de Savoie, CNRS/IN2P3, France,\\
 11: Laboratoire Leprince-Ringuet, Ecole Polytechnique (UMR 7638, CNRS), France,\\
  12: LPNHE, University of Pierre et Marie Curie, Paris 6, University of Denis Diderot, Paris 7, CNRS/IN2P3, France,\\ 
  13: Laboratoire Univers et Particules de Montpellier, Universit\'e de Montpellier, CNRS/IN2P3, France,\\
  14: Grupo de Altas Energias, Universidad Complutense de Madrid., Spain,\\
  15:  Full consortium author list at http://cta-observatory.org
  }\\ 
          {E-mail: \email{glicens@cea.fr}}
}

\abstract{NectarCAM is a camera proposed for the medium-sized telescopes of the Cherenkov Telescope Array (CTA) covering the central energy range of  ~100 GeV to ~30 TeV.  It has a modular design and is based on the NECTAr chip, at the heart of which is a GHz sampling Switched Capacitor Array and a 12-bit Analog to Digital converter. The camera will be equipped with 265 7-photomultiplier modules, covering a field of view  of 8 degrees. Each module includes the photomultiplier bases, high voltage supply, pre-amplifier, trigger, readout and Ethernet transceiver.  The recorded events last between a few nanoseconds and tens of nanoseconds.  The camera trigger will be flexible so as to minimize the read-out dead-time of the NECTAr chips. NectarCAM is designed to sustain a data rate of more than 4 kHz with less than 5\% dead time. The camera concept, the design and tests of the various subcomponents and results of thermal and electrical prototypes are presented. The design includes the mechanical structure, cooling of  the electronics, read-out, clock distribution, slow control, data-acquisition, triggering, monitoring and services.}

\FullConference{The 34th International Cosmic Ray Conference,\\
		30 July- 6 August, 2015\\
		The Hague, The Netherlands}

\begin{document}

\section{Introduction}
\begin{figure}[ht]
\centering
\includegraphics[height=6.5cm]{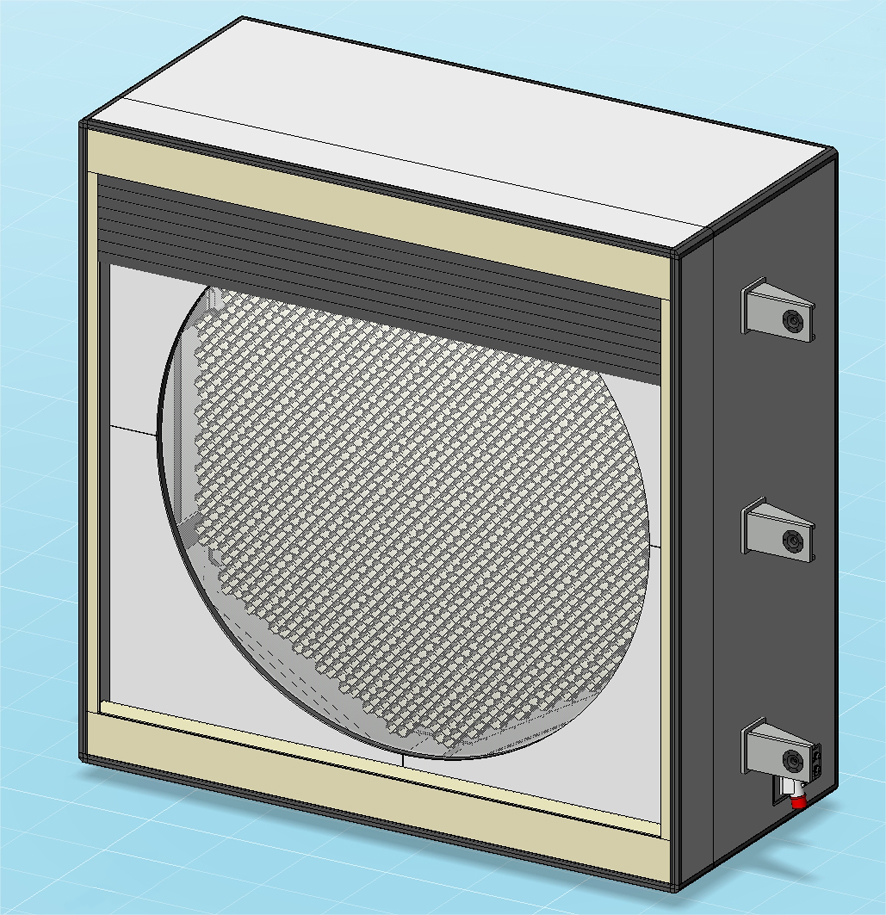}\includegraphics[height=6.5cm]{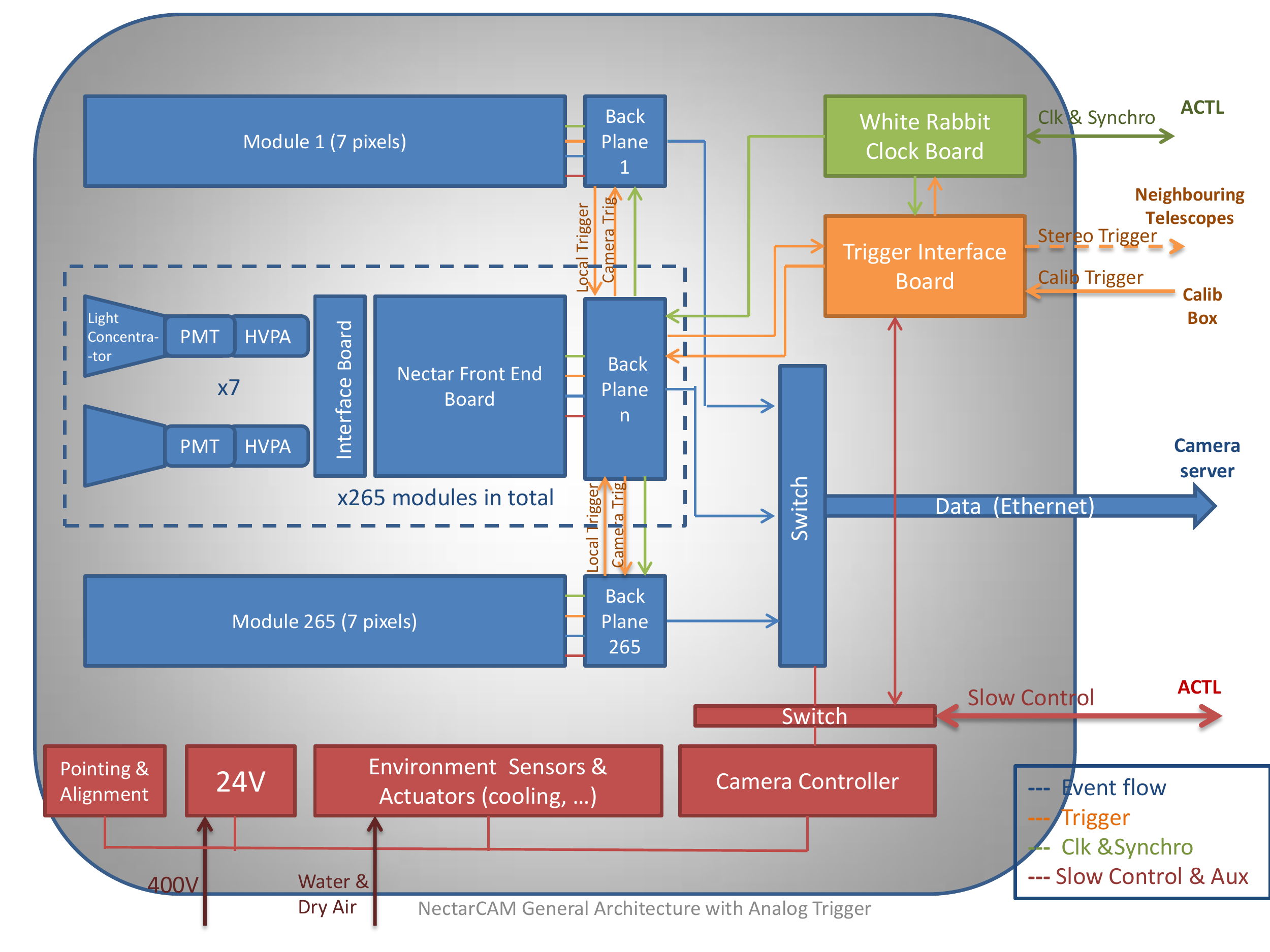}
\caption{Left:  Front view of NectarCAM.  The mounting points on the MST, the shutter and plexiglass window are shown. The module holder can be seen behind the glass window. Right: Global architecture of NectarCAM, with the implementation of the ''analogue trigger'' option.}.
\label{fig:arch-analogue}
\end{figure}

The design of NectarCAM is built on the knowledge gained from the design and operation of H.E.S.S. and MAGIC-II, both of which use analogue memory based cameras. 
It is designed to be installed on the Medium Sized Telescope (MST) of the Cherenkov  Telescope Array (CTA). 
NectarCAM weights 1.9 tons. Its focal plane includes 1855 photomultipliers covering a total field of view of 8 degrees.  It is enclosed in an aluminium box  (LHS of Figure \ref{fig:arch-analogue} and Section \ref{sec:mech}). The reliability of NectarCAM is enhanced by protecting the electronics from outside dirt, by sealing the camera and by carefully controlling the temperature (Section \ref{sec:control}). 
NectarCAM has a modular design. 
The basic element is a module composed of seven photo-detectors with their associated readout and local trigger electronics. The readout is based on a dedicated analogue memory: the NECTAr ASIC (see Reference \cite{6154348} for details). This ASIC has the dual functionality of analogue memory and 12 bit analogue to digital converter. Its memory length is 1024 cells, with a sampling frequency selectable between 0.5 and 2 GHz. The measured analogue bandwidth is more than 250 MHz. The NECTAr chip acts like a circular buffer, which holds the data until a camera trigger occurs.  An Ethernet-based data acquisition system records the full waveform for every pixel. 
The camera is triggered in a multi-step process.  The first step is a local camera trigger.
Two options for the camera trigger are under study: the so-called "analogue" trigger which is an improved version of the H.E.S.S. and MAGIC trigger systems and a novel "digital" trigger. The architectures of the camera are somewhat different depending on the choice of the trigger. RHS of Figure \ref{fig:arch-analogue} shows the architecture of NectarCAM with the implementation of the analogue trigger. 
The photons from the Cherenkov showers are detected by a detector unit, described in Section \ref{sec:fpi}. The resulting signals stored temporarily in an analogue 
memory on the front end boards 
(Section \ref{sec:rdo}).  Trigger conditions are evaluated at any time by a camera trigger system described in Section \ref{sec:trig}. When the trigger conditions are met, the photon signal is digitized, and sent to a camera server (Section \ref{sec:daq}). 

\section{Design}
\subsection{Mechanics and cooling}\label{sec:mech}
The mechanical and thermal sub-systems of NectarCAM are the result of a common design effort with the camera for the large size telescope (LST) of CTA. 
 The mechanics of NectarCAM
is designed to accommodate the modular structure and to allow the easy installation and removal of modules. 
\begin{figure}[ht]
  \centering
  \includegraphics[height=6cm]{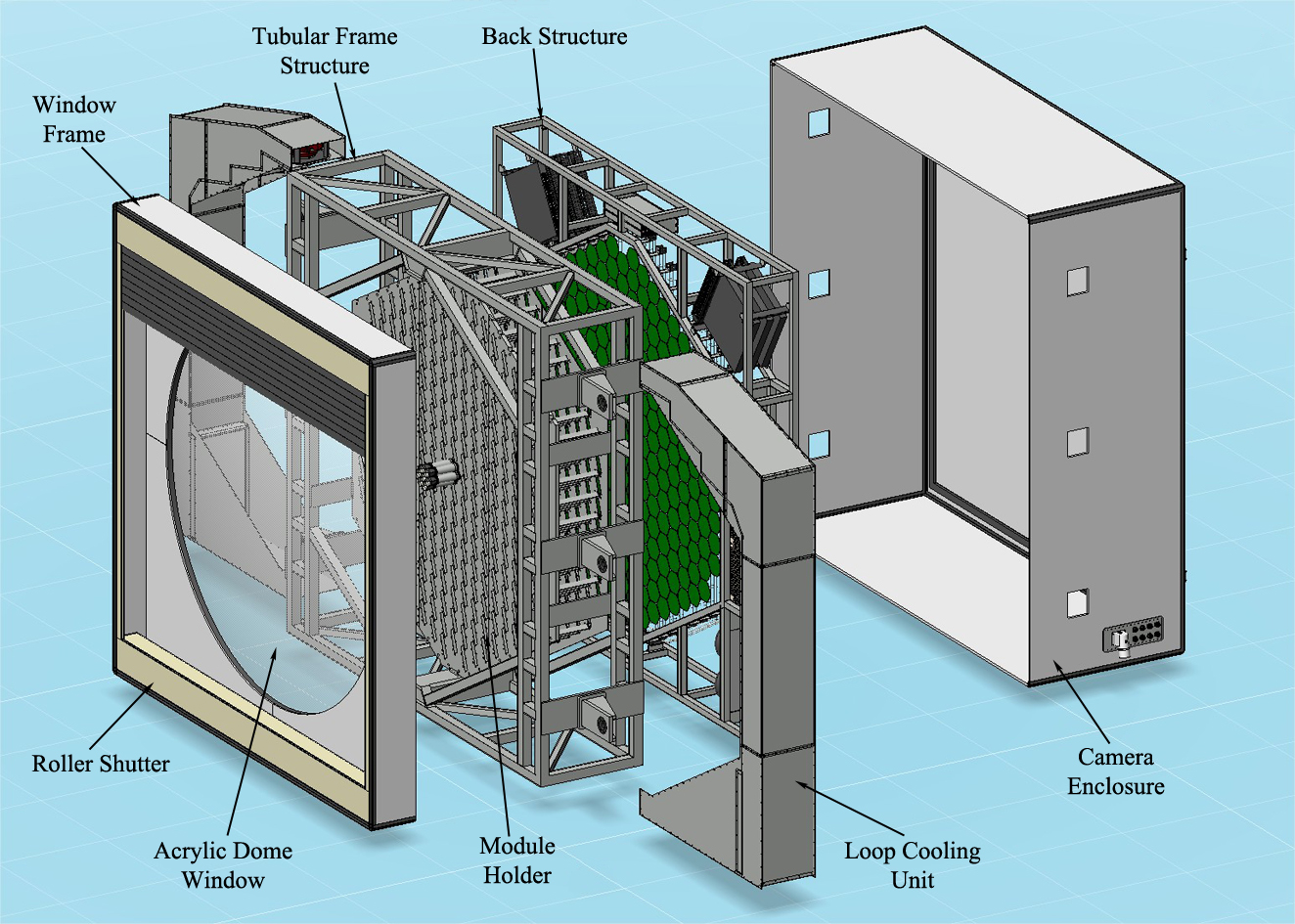}\includegraphics[height=6cm]{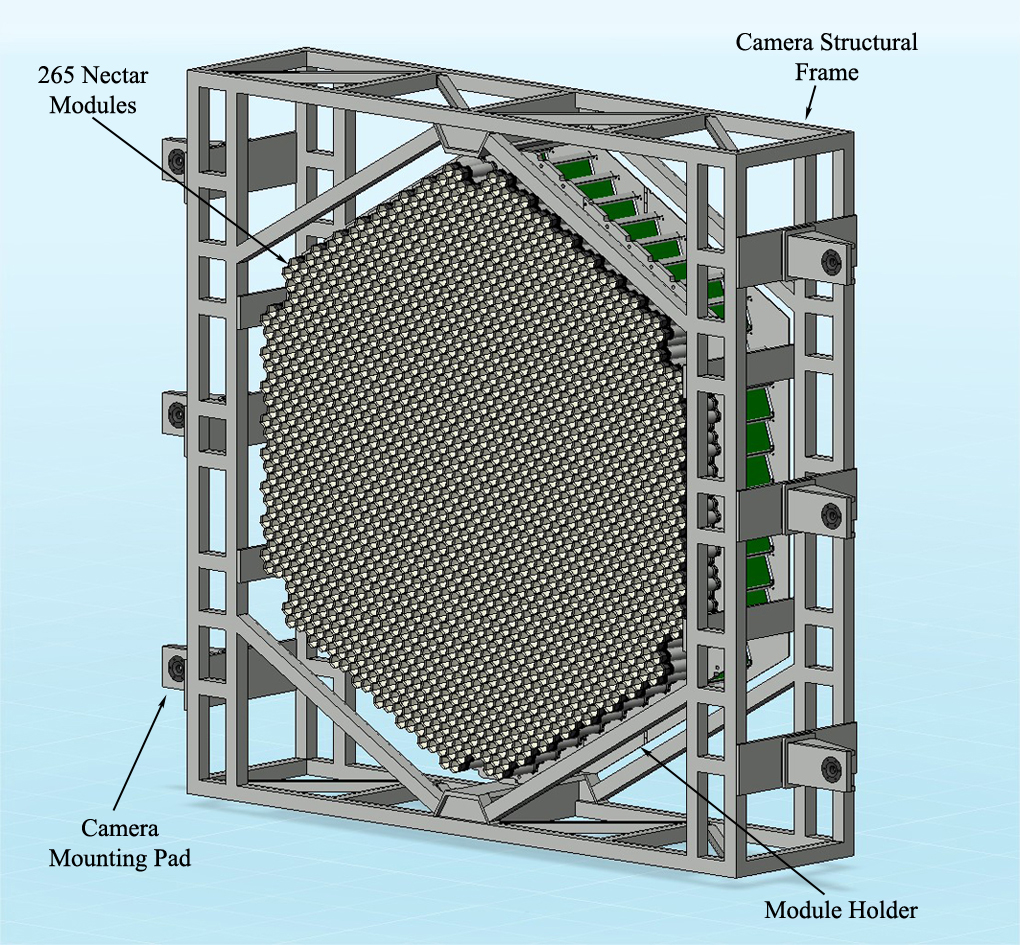}
  \caption{Left: Exploded view of NectarCAM. Right: central part,  module holder.}
  \label{fig:camera_front_exploded}
\end{figure}

NectarCAM is built from several major independent assemblies or units (LHS of Figure \ref{fig:camera_front_exploded}). This architecture provides flexibility to the construction and the integration of the cameras during the production phase.
The front assembly, camera entrance aperture, includes a customized commercial roller shutter to blind the camera and an acrylic sheet that provides the sealing function. Both are mounted on an aluminum frame structure. The front assembly is hinged to the central structure of the camera to provide easy access the modules. 
The central assembly comprises the primary load bearing component of the camera, a tubular frame structure made of welded aluminum profiles, the module holder (RHS of Figure \ref{fig:camera_front_exploded}), the structure that receives all the NECTAr modules, and the cooling units of the frontend electronics. The NectarCAM cooling system has been tested on a prototype, the results of tests are described in Reference \cite{Emmanuel}.   
The back assembly holds the trigger and data acquisition subsystems and all the services of the camera. The devices are mounted on a light self-supporting aluminum structure. 
The camera housing is made from large aluminum honeycomb panels.  


\subsection{Focal plane instrumentation}\label{sec:fpi}
\begin{figure}[ht]
  \centering
  \includegraphics[height=4.8cm]{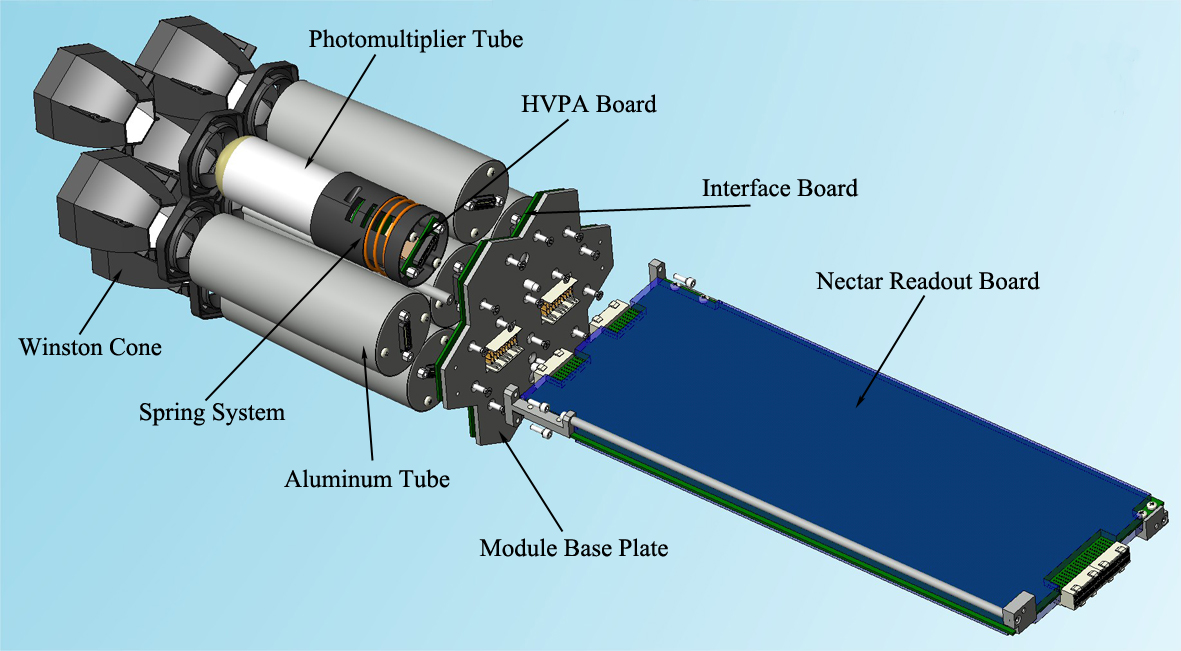}\includegraphics[height=4.8cm]{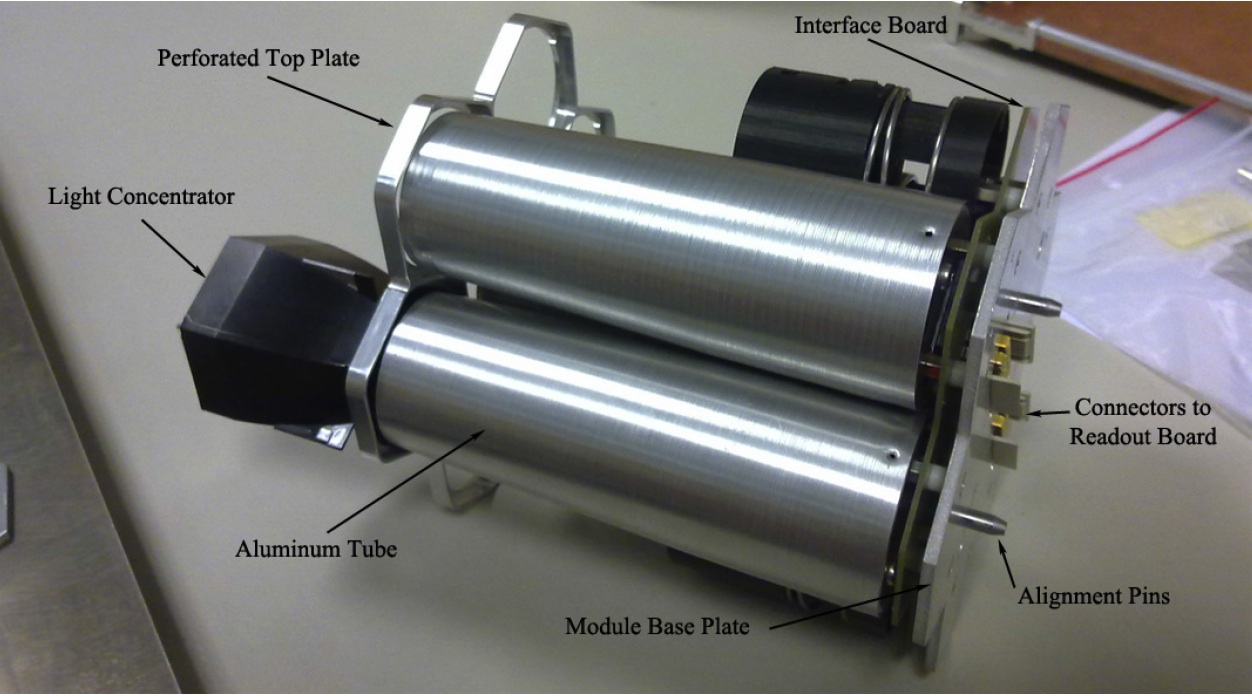}
  \caption{Left: Focal plane instrumentation of NectarCAM, and connection to the front-end board. Right: mechanical model of detector unit.}
  \label{fig:fpi}
\end{figure}

The focal plane instrumentation (FPI) detects the incoming light and converts the photons into an electrical signal. The FPI is located in the front part of the camera body behind the shutter.
The basic building block of the focal plane instrumentation is the association of a high voltage and preamplification board (HVPA) and a phototube into  'Detector Units' shown in Figure \ref{fig:fpi}.  Every phototube is associated to a light concentrator made with hollow cones ('Winston cones'). The cones are made of plastic with a $Al Mg F_{2}$ coating. They are assembled out of  three parts so that only relatively flat surfaces are coated (see Reference \cite{2013SPIE.8834E..05H} for details).
The photomultipler candidate is a customized version of the 1.5'' R12990-100 phototube of Hamamatsu.   The phototubes are operated at a low gain of  $4\cdot 10^4$ to increase their lifetimes. Their signal needs
to be preamplified to compensate for the relatively low gain. 
A HVPA board is soldered to every phototube. It has several functions: creating the high voltage 
of the phototube, setting and monitoring  the high voltage value and finally amplifying the signal to compensate for the low gain of the phototube.  
Finally, the seven Detector Units of a NECTAr module are controlled by a single Interface Board.

\subsection{Read-out and digitization}\label{sec:rdo}
\begin{figure}[ht]
\centering
\includegraphics[height=5cm]{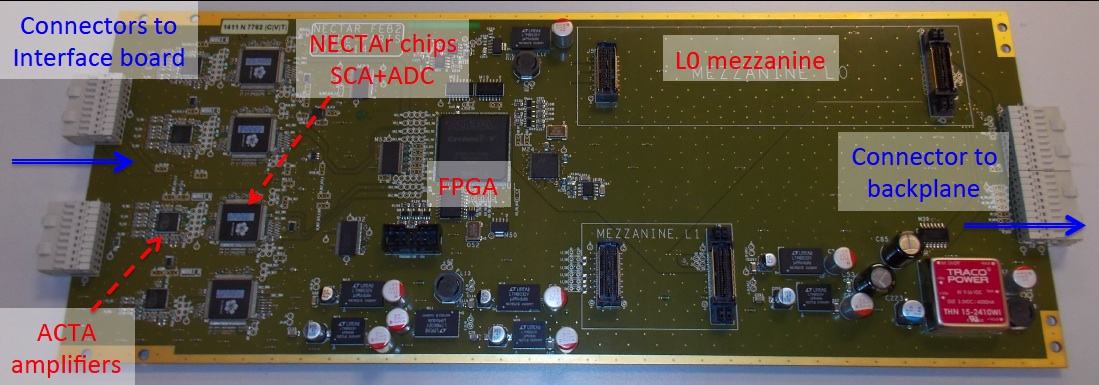}
\caption{Implementation of NectarCAM front-end electronics on the NECTAr V2 prototype. The readout board is equipped with seven NECTAr chips (four on the top of the board and three on the bottom), and has  four  two-channel ACTA amplifiers. Each ACTA amplifier is associated to the output of two photomultipliers. The signal from seven detector units arrives in the left connectors. The digitized signal exits the front-end board by the right connector. The backplane located on the other side of this connector handles the time synchronization signal, the L0 trigger fan-out and the L1 trigger.}
\label{fig:NectarExplained}
\end{figure}

The front-end electronics system captures the signal when a trigger occurs and sends it to the DAQ system via an Ethernet link. 
Its layout is shown in Figure \ref{fig:NectarExplained}. 
The dynamic range of the output signal
is 0 to 2000 photo-electrons, with the capacity of measuring the single photoelectron at the nominal photomultiplier high voltage, to allow a simple calibration.  The full dynamic is covered by two electronics channels with different gains. 
   
Each front-end board reads out the signal from 7 phototubes.  Front-end boards include:
\begin{itemize}
\item ACTA amplifiers. These custom made amplifiers have outputs to two data channels (low and high gain) and a trigger channel for two photomultipliers. They were designed to have a low power consumption while keeping a large enough bandwidth.
\item NECTAr chips which are used for storing and digitizing the photon signals with a 11.3 bit effective accuracy. 
\item Front-end-board management and interface to Ethernet.  
\item The trigger L0 and optional L1 ASICs (see Section \ref{sec:trig}), used with the ''analogue trigger'' option. 
\end{itemize}  
The front-end board is plugged to a 'backplane board' which is described in Section \ref{sec:trig}.

\begin{figure}[h]
\centering
\includegraphics[height= 6cm]{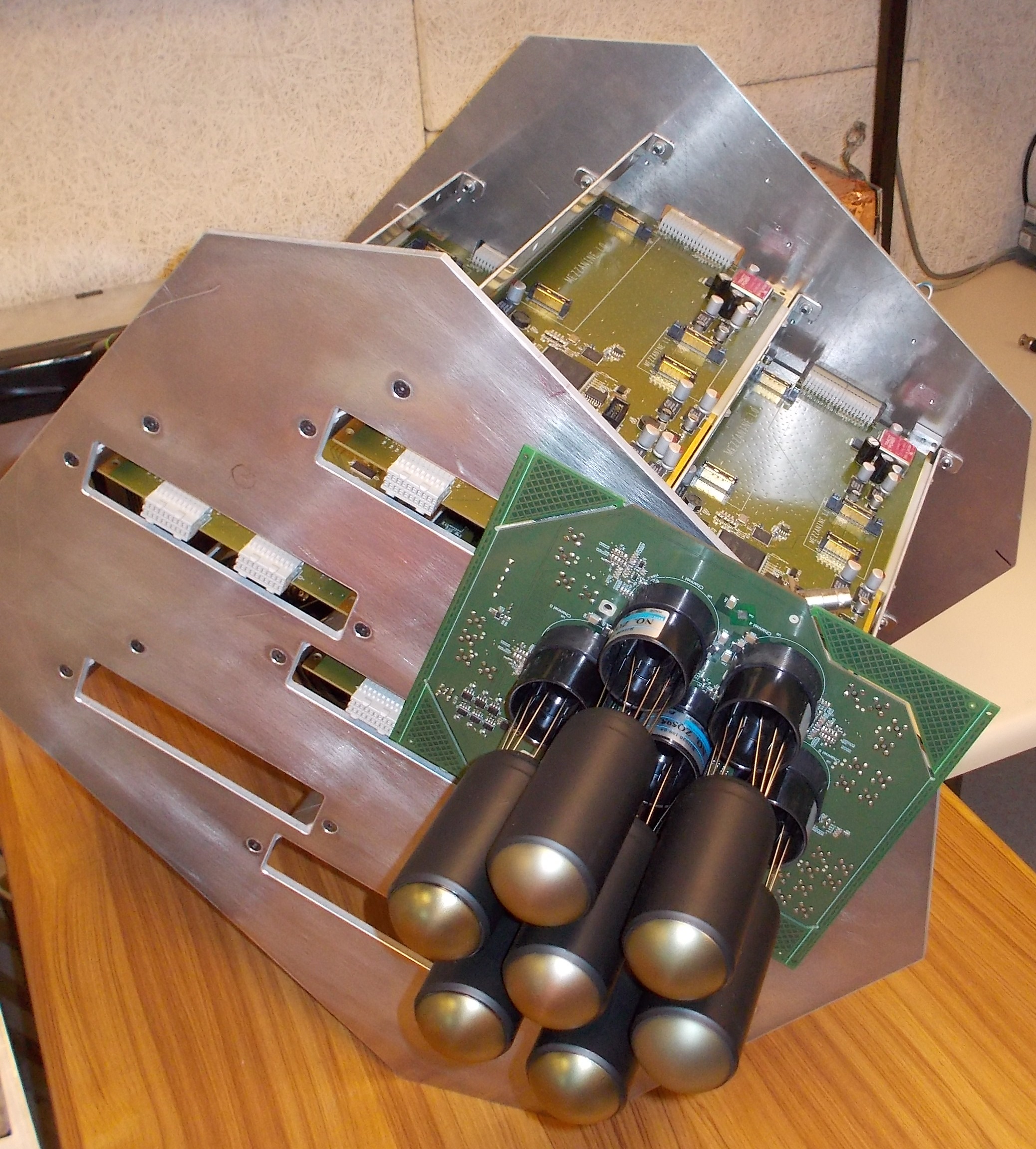}\includegraphics[height= 6cm]{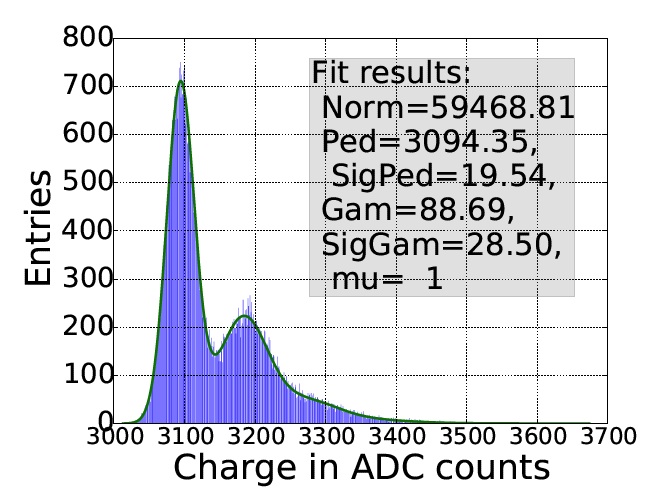}
\caption{Left: Test equipment for the single module test bench. Right: Single photo-electron measurement.}
\label{fig:singlepe}
\end{figure}

\begin{figure}[h]
\centering
\includegraphics[width= 9.5 cm]{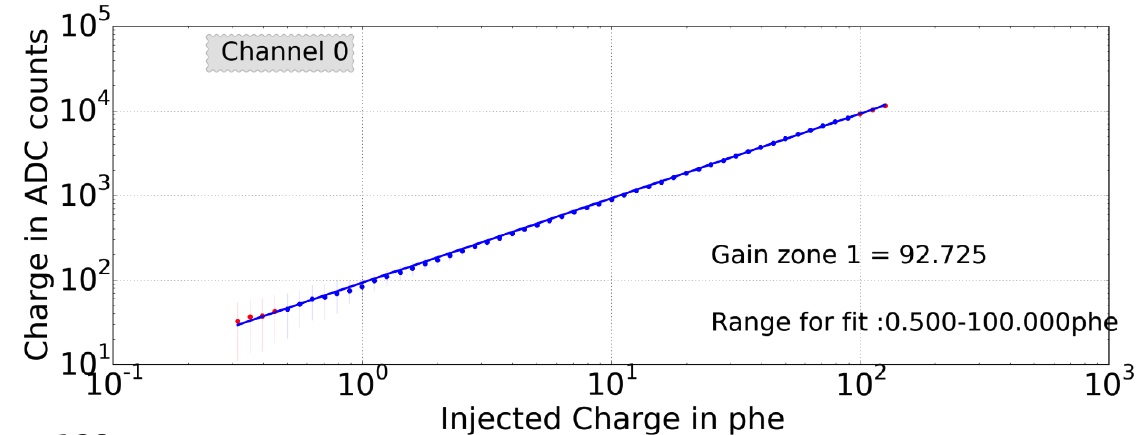}\includegraphics[width= 6.5 cm]{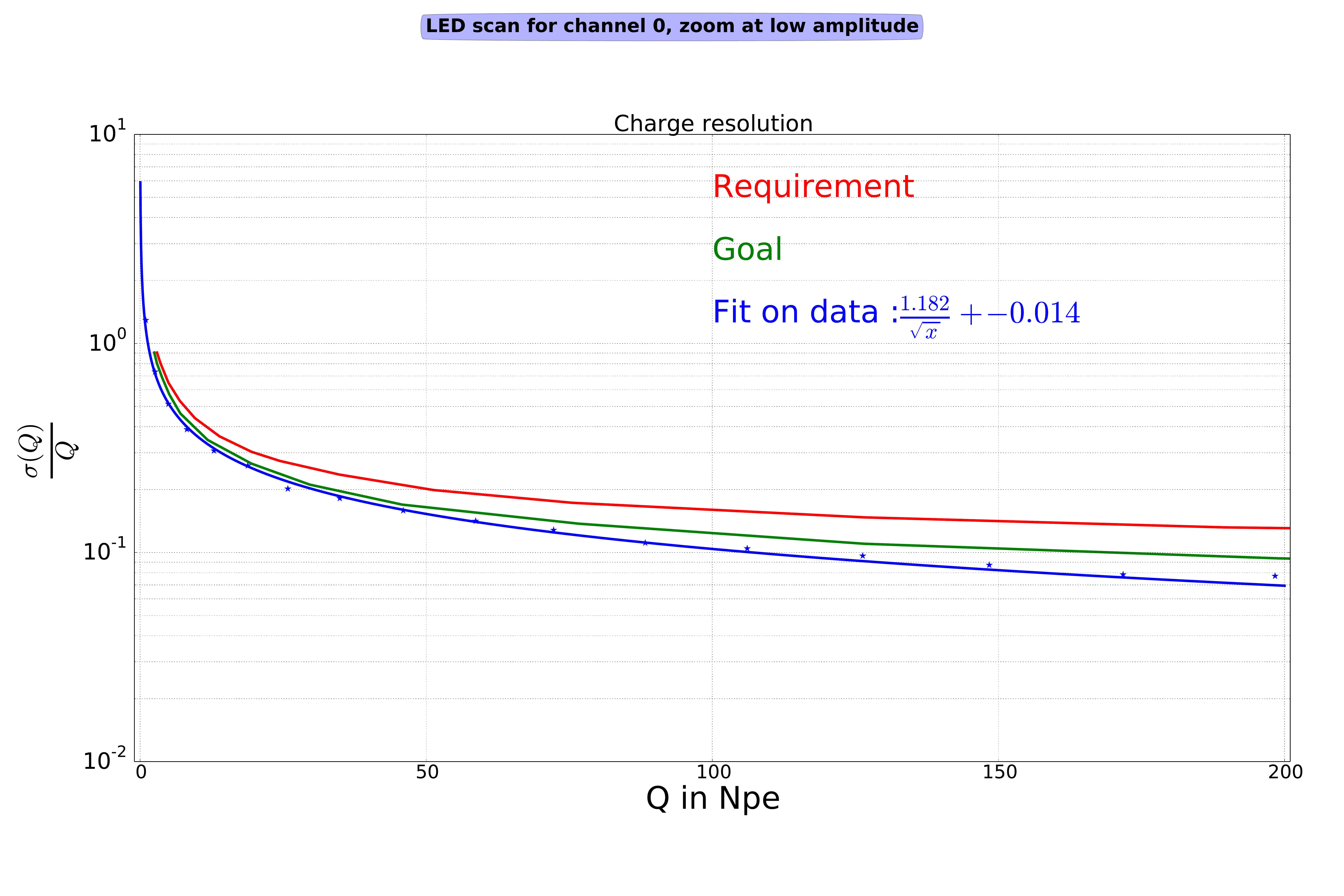}
\caption{Left: Linearity measurement on a high gain channel. Right: Charge resolution versus injected charge (high gain channel).}
\label{fig:FEBlinearity}
\end{figure}

The performance of the front-end electronics have been extensively studied with electrical and photon signals. The LHS of Figure \ref{fig:singlepe} shows Nectar modules equipped for the single module test bench. 
The single photon spectrum is obtained by using a LED pulser at a low light level and fitting the positions of multiple photoelectron peaks. The gain of the photomultiplier is set to the nominal value  of  $40000.$   The read-out and integration window is 16 ns. An example is given on the RHS of Figure \ref{fig:singlepe}, where mainly the 1-pe peak is visible.  The light input had an average of 1 photon per trigger. The measured single photo-electron charge resolution is 32\%. 
The analogue chain of NectarCAM is linear to better than 5\% between 0.5 and 2000 photoelectrons, as illustrated on the LHS of Figure \ref{fig:FEBlinearity} for the high gain channel.  
The corresponding charge resolution is shown in the RHS of Figure \ref{fig:FEBlinearity}. 

\subsection{Trigger}\label{sec:trig}
NectarCAM is triggered when it receives a significant amount of light in a compact region of the focal plane (of the order 0.2 deg$^{2}$) within a short period of time (typically a few ns). Translated in terms of camera hardware, 
this means that several neighbouring modules  will see photon signals within a few ns. The basic strategy is thus to trigger first a module or a neighboring group of pixels (level 0 trigger or L0) and then the camera (Level 1 or L1 and possibly Level 2  or L2).  The final decision is taken by the array level trigger.   
NectarCAM is expected to have a fully efficient trigger above $\sim 80$ GeV.

NectarCAM is evaluating two different camera trigger system designs. The so-called 'Analogue Trigger' (see Reference \cite{2013ITNS...60.2367T} for details), which leads to the architecture of the RHS of Figure  \ref{fig:arch-analogue}, is based on the analogue summation of the signal of neighbouring pixels at L0, along with summation of L0 signals at camera trigger level (L1). This trigger scheme is known to be very efficient at low energies 
and will be installed on the LST camera.  The other possibility, (see Reference \cite{Shayduk})  is the 'Digital trigger' which uses combinations of digitized input at L0. This fully digital trigger scheme is not as efficient for triggering at low energies but is more flexible.

In the analogue trigger scheme, at trigger level L0, pixel signals from a
single module are summed, and the resulting L0 signal distributed to the neighboring modules to perform additional sums. This final sum of neighboring L0 signals is compared against a threshold to obtain a camera 
trigger decision (L1). In a similar fashion, in the Digital Trigger, the discriminated L0 pixel signals are digitally combined in its own L1 backplane to   produce the camera trigger decision (L1). 
Once the L1 trigger decision has been locally taken, it can be distributed either to neighbouring modules or to the whole camera. 
There are common components to both the Analogue Trigger and the Digital Trigger design, namely the Trigger Interface Board and the L0 ASIC. 

\begin{figure}[h]
\centering
\includegraphics[width=9cm]{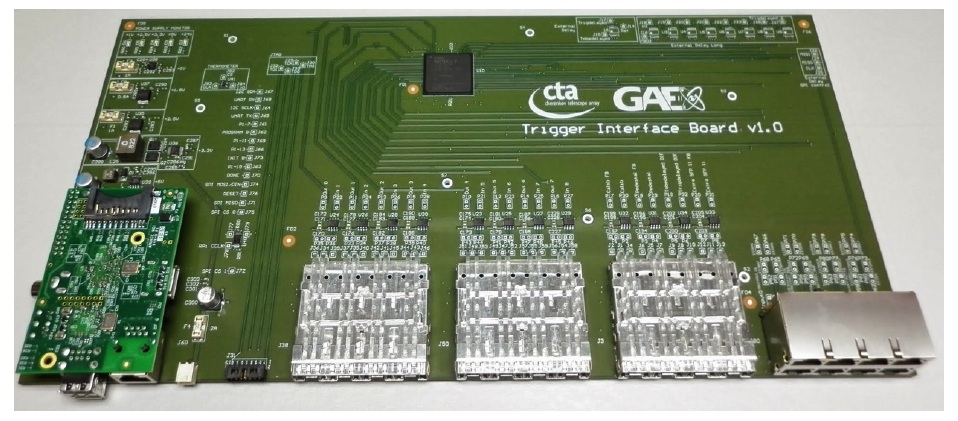}\includegraphics[width= 4cm]{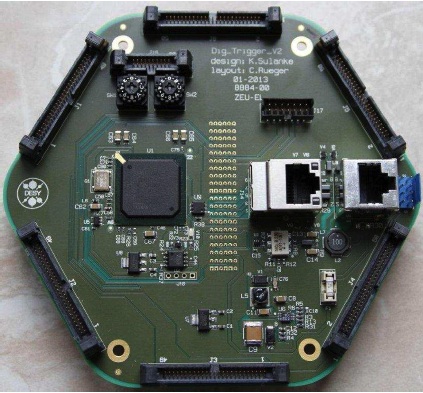}
\caption{Left: Trigger interface board prototype. Right: Digital backplane board.}
\label{fig:atb}
\end{figure}

The trigger interface board (LHS of Figure \ref{fig:atb}) performs two basic operations.
The most important of these operations is to provide an interface between the camera trigger and the rest of the relevant systems of CTA, such as the Camera Clock Board, based on the White Rabbit\footnote{http://www.ohwr.org/projects/white-rabbit} standard or the different calibration signals generated by the telescope or the observatory itself. The Trigger Interface board distributes signals such as the L1 Accept, or the Trigger Type to the modules. As an extra,  it will allow forming coincidences among camera trigger signals of neighbouring telescopes to generate a local stereo trigger. 

The L0 ASIC processes analog signals from the individual pixels in a module. The signal from the pixels are first sent to analogue delay lines, which compensate for the different time shifts introduced by phototubes with different high voltage settings. The pixel signals are treated and then added together before being sent to the Level 1 decision trigger. Two different Level-0 approaches have been included in the L0 ASIC: the Majority trigger and the Sum trigger. 
The Majority trigger concept compares the signal from each pixel to a voltage threshold (also programmable) using a discriminator circuit. Each differential output pair of the discriminator is available as a LVDS digital output. 

Both analogue and digital options 
implement part of their functionalities in dedicated Back-Plane Board (respectively called Analog Trigger Back-Plane and Digital Trigger Back-Plane), which have the same interface to the front-end board, that allows testing both systems with the camera demontrators.
The backplane board is the only electrical interface of the front-end board to the outside world, so it
must deliver all the additional servicing required for normal
operation of the module. Thus, in addition to the trigger functionalities, it provides the front-end
board with Ethernet (for data transfer) and power interfaces,
as well as IP address.  Finally, the backplane board
receives from the L1 distribution system the synchronization signals required by the front-end electronics.
\subsection{Data Acquisition (DAQ)}\label{sec:daq}
The design of the NectarCAM DAQ is based on the use of standard, off-the-shelf networking and computing hardware components, with a  reasonable time-on-market. It is described
in detail in Reference \cite{Dirk} to which the reader should refer for more details. The prototyping effort has been concentrated on the two most critical items: the camera server and its software model  and the Ethernet switches.
\subsection{Slow control}\label{sec:control}
\begin{figure}
  \centering
  \includegraphics[width=7cm]{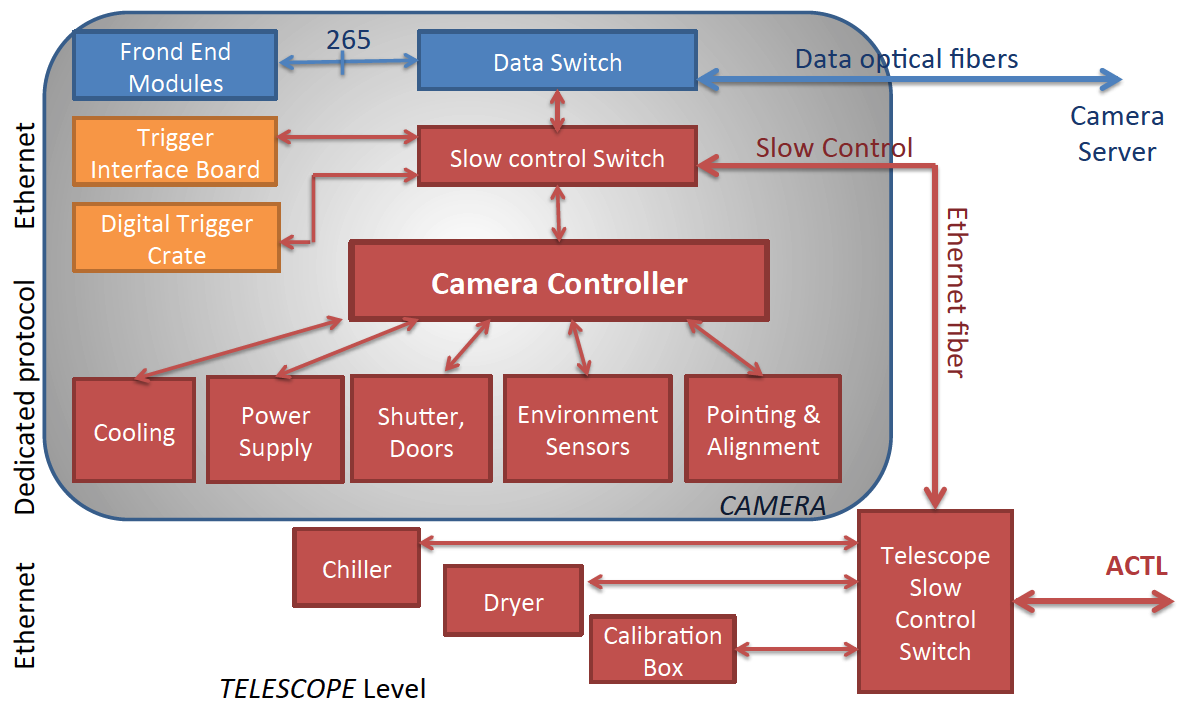} \includegraphics[width=5cm]{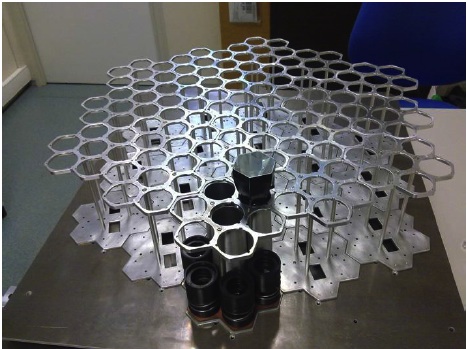}
  \caption{Left: Functional architecture of the slow control of NectarCAM. Right: module holder for the 19-module mini-camera. }
  \label{fig:slc-arch}
\end{figure}

The slow control system of NectarCAM controls the camera and ensures its safety. Its functional architecture is shown in Figure \ref{fig:slc-arch}. The first level safety is enforced by the Embedded Camera Controller, a compactRIO controller from National Instrument with a real time operating system, which handles the interfaces with most of the slow control subsystems. The Embedded Camera Controller has an OPC-UA interface for the remote control by the Array Software. 

\section{Conclusion and plans}
The NectarCAM development plan is divided into three distinct phases. The present phase (pre-construction, until the end of 2016) aims at building prototypes to validate design choices.
These prototypes included a single module, for the validation of the detection performance, then a 5-module cluster, for the validation of the interface between modules and backplanes, a thermal prototype and finally a 19-module mini-camera. This 19-module mini-camera (see RHS of Figure \ref{fig:slc-arch}) will validate the operations of NectarCAM and
help decide between the analogue and digital trigger options.   
The second phase (pre-production) aims at assembling a full camera qualification model and producing the test benches and tools for the assembly to be used for mass production.
The last phase, starting in 2018 is hoped to be the production and delivery of 23 cameras, including the spares, to the sites of CTA.

\section*{Acknowledgements}
This work has been carried out thanks to the support of the OCEVU Labex (ANR-11-LABX-0060) and the A$\star$MIDEX project (ANR-11-IDEX-0001-02) funded by the "Investissements d'Avenir" French government program managed by the ANR. 
We gratefully acknowledge support from the agencies and organizations listed
on  Web page http://www.cta-observatory.org/?q=node/22.

\end{document}